# Magnetodielectric coupling in a Ru-based 6H-perovskite, Ba$_3$NdRu$_2$O$_9$

*Tathamay Basu,[1,2,*] Alain Pautrat,[1] Vincent Hardy[1] Alois Loidl,[2] Stephan Krohns,[2]*

[1]*Laboratoire CRISMAT, UMR 6508 du CNRS et de l'Ensicaen, 6 Bd Marechal Juin, 14050 Caen, France*

[2] *Experimental Physics V, Center for Electronic Correlations and Magnetism, University of Augsburg, Universitätsstr. 2, D-86135 Augsburg, Germany*

[*]*tathamaybasu@gmail.com*

## Abstract

**A large spin-orbit coupling is a way to control strong magnetodielectric (MD) coupling in a higher *d*-orbital materials. However reports are rare on such compounds due to often leaky conductive behavior. Here, we demonstrate MD coupling in a Ru-based 6H-perovskite system, Ba$_3$NdRu$_2$O$_9$. The rare-earth ion in a 6H-perovskite makes the system insulating enough to carry out MD investigation. The compound is ferromagnetically ordered below 24 K ($T_C$), followed by another magnetic feature at T~ 17 K ($T_2$). The dielectric constant clearly traces the magnetic ordering, manifesting a peak at the onset of $T_C$, which is suppressed by the application of an external magnetic field (*H*). The results indicate the presence of MD coupling in this compound, which is further confirmed by the *H*-dependence of the dielectric constant. Interestingly, a cross-over of the sign of MD coupling is observed at T ~ $T_2$. We conclude that two different mechanism controls the MD coupling which yields positive and negative coupling, respectively. Both mechanisms are competing as a function of temperature and magnetic field. This brings us a step closer to design and control the magnetodielectric effect in 6H-perovskites containing higher *d*-orbital elements.**

The research on higher *d*-orbital materials (4*d*/5*d*) has paid significant attention due to its exotic magnetic behavior as a result of various competing effects like extended orbitals, pronounced crystal-field effects and strong spin-orbit coupling.[1] The higher *d*-orbital is also of interest in the field of magneto(di)electric coupling (MD) and multiferroicity due to the possibility of large MD coupling originating from the enhanced spin-orbit coupling. Systems, crystallizing in perovskite structure exhibiting magnetic frustration and containing rare-earth (R) and 3d metal ions (such as RMnO$_3$, R=Ho, Tb, etc.), are vastly explored in the field of multiferroicity and MD coupling.[2] It has been theoretically predicted that 3*d*-5*d* double perovskite system, such as Bi$_2$MReO$_6$ (M=3d metal) and Zn$_2$FeOsO$_6$ could serve as good multiferroic and magnetoelectric system as a result of higher *d*-orbital.[3,4] The investigation through non-contact spectroscopic method also supports this framework.[5] However, experimental reports of MD effects in magnetic 4*d* or 5*d* materials are rare in literature,[6] because these systems are less insulating due to the large extend of the *d*-orbitals leading to leaky conductive behavior. This hampers the investigation of bulk dielectric/ferroelectric properties experimentally. The compound Co$_4$Nb$_2$O$_9$ exhibits giant MD coupling,[7] though, Nb is non-magnetic here and it may not has any direct role on MD coupling. However, no attention has been paid to multiferroicity or/and MD coupling in the 6H-perovskite system containing higher *d*-orbital (magnetic) ion, such as, Ba$_3$MRu$_2$O$_9$ (M=3d-ion, Bi, Y, La, R, etc.).[8,9,10,11,12] It is experimentally observed that introduction of rare-earth ion (in place of Bi/La/Y) makes the system more insulating due to the localized atomic nature of rare-earth. For an example, Haldane spin-chain system Y$_2$BaNiO$_5$ is not highly insulating and does not show any MD coupling or multiferroicity,[13] but R$_2$BaNiO$_5$ reveals highly insulating behavior and exhibits multiferroicity with strong MD coupling.[14] Also, *d-f* magnetic correlation plays a significant role in the complex magnetism and thus could have an interesting effect on MD coupling.

We show that Ba$_3$NdRu$_2$O$_9$ is a promising system revealing highly insulating behavior even at temperature below magnetic ordering, allowing the investigation of its dielectric properties. We have performed dc and ac magnetization, dielectric and magnetodielectric measurements as a function of temperature and magnetic field. There is no earlier report of multiferroicity/magneto(di)electric coupling in this $Ba_3RRu_2O_9$ system, or even in a Ru-based system.

The system with general formula $Ba_3MRu_2O_9$ crystallizes in 6H-perovskite structure and consists of $Ru_2O_9$ dimer (face-sharing distorted $RuO_6$ octahedral) and regular corner sharing $MO_6$ octahedral.[8,9] Depending on M-ion, the system $Ba_3MRu_2O_9$ exhibits different magnetic behavior; for M= non-magnetic ion, it behaves like a dimer ($Ru_2O_9$) system;[15] and for magnetic ion (M= $Ni^{+2}$/ $Co^{+2}$), another magnetic super-exchange path Ru-O-M is established and long-range magnetic ordering is developed.[6] If M is replaced by lanthanides ($Y^{+3}$, $La^{+3}$, $R^{+3}$), the system can crystallize in the 6H-perovskite structure and charge neutrality is balanced by changing the effective valance state of Ru (+4.5, one Ru-site is $Ru^{4+}$ and the other site is $Ru^{+5}$ in $Ru_2O_9$ dimer).[9,10] This mechanism is confirmed by neutron diffraction as well.[11] Exceptionally, only in the case of $Ba_3Bi(Ru/Ir)_2O_9$, Bi possesses an unusual valance state of +2, rather than +3 and Ru has usual +5 state.[16] Interestingly, the compound containing M=3d-metal ion/Bi is semiconducting in nature, whereas, the introduction of R-ions makes the system a rather good insulator. The title compound $Ba_3NdRu_2O_9$ is ferromagnetically ordered below 24 K ($T_C$), followed by another complex magnetic feature at ~17 K ($T_2$), which is attributed to a canted antiferromagnetic (AFM) ordering.[10,11] Magnetic frustration was already predicted in this compound.[11] Hence, this is a good system to explore MD coupling.

The compound $Ba_3NdRu_2O_9$ is synthesized by solid-state-reaction and forms a single phase with expected P63/mmc space group, as reported earlier.[10] The temperature ($T$) and magnetic field ($H$) dependent dc and ac magnetic susceptibility ($\chi$) measurements are performed using a Superconducting Quantum Interference Device (SQUID, Quantum Design). The complex dielectric measurement has been carried out as a function of temperature and magnetic field with a 1 V ac bias for various frequencies using a LCR meter (Agilent 4284A) with a home-made sample insert for a Physical Properties Measurement System (PPMS, Quantum Design). Silver paint is used to make parallel plate capacitor of a pressed disc-like polycrystalline sample (5 mm diameter and 0.9 mm thickness).

The preliminary dc magnetic results were already reported by Doi, et al.,[10] however, here we document detailed dc magnetization elucidating the magnetic behavior and for comparison to check the quality of sample (See Supplementary Material for detail results and discussion). In the present manuscript, we focus on the magnetic behavior in the presence of high magnetic fields and far above the ordering temperature $T_C$, which was not reported so far. Further ac susceptibility measurements in presence of different dc magnetic fields are performed. The real ($\chi'$) and imaginary ($\chi''$) part of the ac susceptibility is shown in figure 1, as a function of temperature ($T$) for a fixed frequency of 111 Hz and in the presence of different dc magnetic field ($H_{dc}$). The distinct peak anomaly in zero dc magnetic field ($H_{dc}$) at $T_C$ in both $\chi'(T)$ and $\chi''(T)$ confirms the magnetic ordering (figure 1a and 1d). The application an external dc field, $H_{dc}$, decreases the magnitude of $\chi'(T)$ and the peak becomes broader and shifts towards higher temperature with increasing $H_{dc}$ (figure 1a, 1b and 1c for $H_{dc}$ =0, 10 and 50 kOe, respectively). Obviously, $\chi''(T)$ is almost zero for high $H_{dc}$ (figure 1e and 1f). Generally, $Tc$ increases slightly with increasing dc field in a ferromagnetic system. However, such a large shift of peak temperature in susceptibility in presence of $H_{dc}$ (that is, 24 K to nearly 30 K from H=0 to 50 kOe in $\chi'(T)$) is not at all typical for a ferromagnetic system. These results are in good agreement with dc magnetization, where the inverse susceptibility ($\chi^{-1}(T)$) does not superimpose for different $H$ far above $T_C$ (see Supplementary Material for detail discussions). In addition the isothermal magnetization exhibits a non-linear behavior even at 30 K (see Supplementary Material for detail). The reported results suggest the presence of magnetic correlation far above $T_C$ and point towards complex magnetic interactions in this compound. Similar complex magnetic behavior is demonstrated in a geometrically frustrated spin-chain system $Ca_3Co_2O_6$.[17]

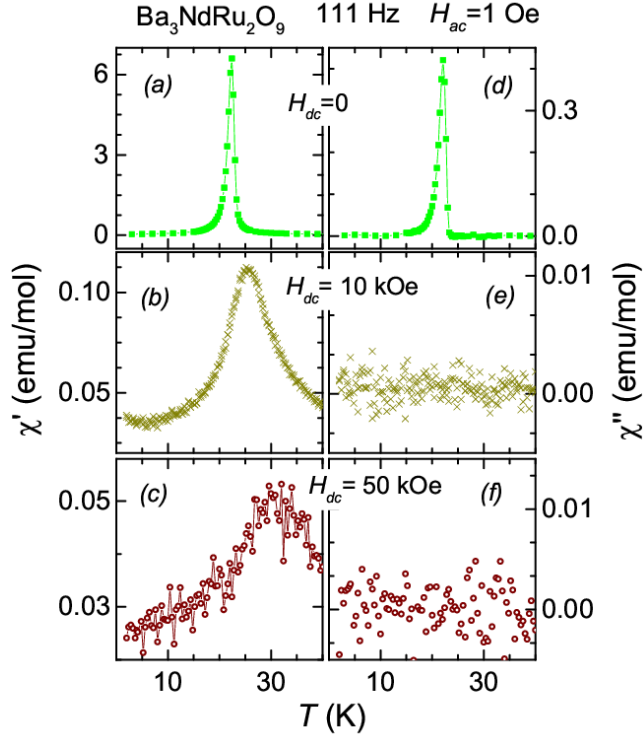

**Figure 1.** Real (left panel) and imaginary (right panel) part of ac magnetic susceptibility as a function of temperature for zero dc magnetic field (a) as well as $H_{dc}$= 10 kOe (b) and 50 kOe (c).

The temperature dependent dielectric constant is depicted in figure 2 for 51 kHz, for two different orientation of $E$ and $H$, that is, $E//H$ (figure 2a) and $E{\perp}H$ (figure 2b). No frequency dependence dielectric feature is observed at low temperatures in the ordered magnetic state (see Supplementary Material). The value of the loss angle (tan$\delta$) as a measure of the dielectric loss is rather low (<0.1), as shown in the inset of figure 2a and 2b confirming the insulating nature of the sample. These dielectric properties allow investigating MD coupling in the magnetic ordered state. The real part of dielectric constant ($\varepsilon'$) clearly exhibits a peak at the onset of $T_C$ in absence of a magnetic field. Interestingly, the applications of a magnetic field strongly influence the dielectric constant in the magnetic regime. The dielectric peak at the ordering temperature $T_C$ gradually suppress with gradual increase of magnetic fields for both the orientations, $E//H$ (figure 2a) and $E{\perp}H$ (figure 2b). For $E{\perp}H$, the dielectric constant decreases with decreasing temperature and its value is lower for higher magnetic fields throughout the temperatures investigated. However, for $E//H$ orientation, a cross-over of the dielectric constant for different magnetic fields (with respect to zero magnetic field) is observed at T~ 13-16 K. The cross-over point (denoted by the dashed line in figure 2a) slightly varies with different magnetic fields. The value of $\varepsilon'$ gradually becomes reduced with the increase of $H$ for $T>16$ K for all H≥10 kOe and the value of $\varepsilon'$ increases with increasing $H$ for $T<13$ K for H>10 kOe, as depicted in figure 2a. A negligible change in the value of $\varepsilon'$ is observed for H≤10 kOe.

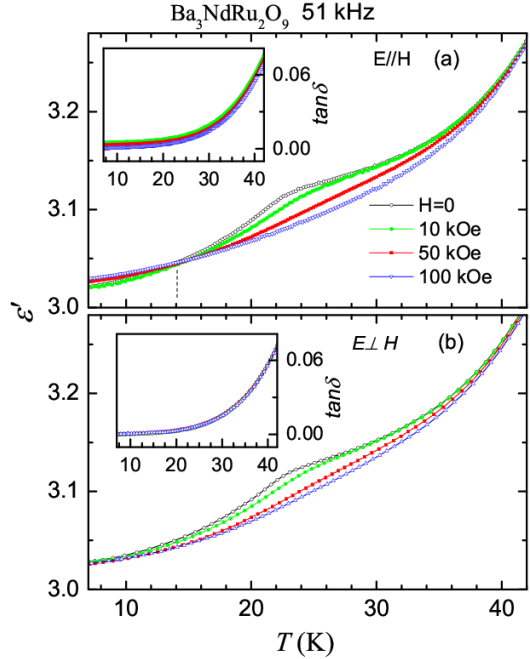

**Figure 2. Real part of dielectric constant as a function of temperature for different magnetic field for (a)** *E//H* **and (b)** *E⊥H* **orientation. Insets show the loss tangent, respectively.**

We have further confirmed the MD behavior through measurements of the *H*-dependent dielectric constant, as shown in figure 3a (*E//H*) and 3b (*E⊥H*) for a fixed frequency of 51 kHz. The features are very similar for others frequencies (not shown here). For *E//H* orientation (see figure 3a), a quadratic positive MD is observed at low *T* (for *T*≤12 K), however, at *T*=15 K, the MD initially decreases with increasing *H* up to 25 kOe, become nearly flat before it increases again with increasing *H* above 40 kOe. Therefore, there is a cross-over of MD sign from negative to positive with increasing *H*. At *T*=20 K, negative MD is observed (figure 3a). Therefore, two components must exist, which are varying different as function of *T* and *H* and probably are quite complex. The positive MD component completely dominates at low *T* (and high *H*), but the negative MD component competes with increasing *T* and dominates at further increasing temperatures (say, 20 K). For the *E⊥H* orientation, negative MD nearly dominates at all temperatures (figure 3b), consistent with the *T*-dependent results in the presence of magnetic fields. However, positive MD term exists at low temperatures as well for *E⊥H* (see inset of figure 3b for 8K). Therefore, both, positive and negative component of MD, are present for the orientations of *E* and *H*, but only the absolute values of MD differ for these two orientations. This signals anisotropy effects within experimental uncertainties. Anisotropic MD behavior for different orientation of *E* and *H* (*E//H* and *E⊥H*) in polycrystalline form is rare but earlier reported in a hybrid metal-organic system.[18] Appreciable negative MD is observed above $T_C$ for both the orientations of *E//H* and *E⊥H* (e.g. 30 K and 40 K) as shown in figure 3a and 3b, which infers the existence of MD even above long-range magnetic ordering due to presence of magnetic correlation. The small MD far above long-range ordering (even from short-range magnetic interaction) is observed in many oxide compounds exhibiting complex magnetic ground states (e.g., Ref.[17,19,20,21]). However, the sample becomes less insulating with increasing temperature T≥25 K (loss part tanδ increases with *T* above 25 K, inset of figure 2), hence, the presence of a very small leakage current above 25 K cannot be ruled out. The magnitude of small negative MD at T≥25 K may be amplified further by the leakage contribution of Maxwell–Wagner-like origin,[22] even in the case that the dielectric feature is of intrinsic origin. However, the sample is highly insulating below $T_C$ (very low tanδ value which is nearly constant throughout $T$-range below $T_C$). The strong positive MD below 15 K and the change of the nature (i.e., sign) of MD with $H$ and $T$ cannot originate from leakage currents.

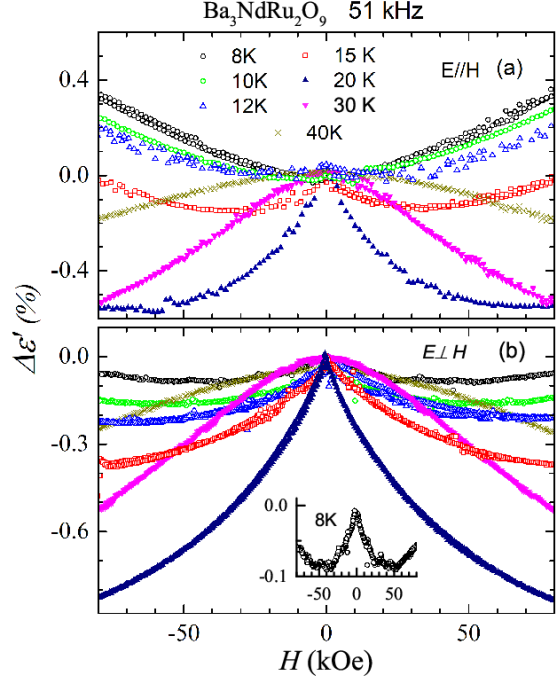

**Figure 3.** Fractional change in $\varepsilon'$ ($\Delta\varepsilon' = [\varepsilon'(H) - \varepsilon'(0)] / \varepsilon'(0)$) as a function of magnetic field for a fixed frequency of 51 kHz at selected temperatures for (a) $E//H$ and (b) $E\perp H$. Inset of (b) shows $\Delta\varepsilon'(H)$ at 8K for $E\perp H$.

Systems containing dimers often exhibit magnetoelastic coupling, which is reported in this series as well.[12] Therefore, the small negative MD component may arise via magnetoelastic coupling. The presence of MD (absolute value could be amplified by leakage current) above $T_c$ is also consistent with this conclusion. Strong positive MD shows up below ~ $T_2$. The canted antiferromagnetic ordering of Nd moments and AFM ordering of ferromagnetic dimer is established below $T_2$.[11] Often Dzyaloshinskii-Moriya (D-M) interaction establishes in weak ferromagnetic or canted AFM systems, which may govern MD coupling. The magnetic frustration could be minimized via lattice distortion to a stabilized ground state at low temperature or/and high magnetic fields, which in turn may favors positive MD coupling. Hence, magnetic fields have a pronounced effect on the magnetic ordering as well as the MD coupling in the present sample. Neutron diffraction investigation by Senn *et al.*[11] demonstrates the thermal evolution of the magnetic moment of this compound revealing that Nd-moment (z-component $m_z$) start to order below 25 K (along '*c*'-direction) and the moment ($m_z$) sharply increases with decreasing temperature nearly saturating below 15 K. In contrast, the other two magnetic components ($m_x$ and $m_y$) in 'ab'-plane from both Nd and Ru moments emerges below 18 K and slowly saturate below the same temperature of about 15 K. Interestingly, the MD cross-over (i.e., emergence/dominance of positive MD component) as a result of competing positive and negative MD occurs nearly at the same temperature (~15 K). The feature and change in the value of isothermal MD is nearly the same below 15 K (see figure 3a for T=8, 10 and 12 K), where magnetic ground state is fully stabilized and moments are almost saturated. The clear lattice anomalies at the onset of two magnetic transitions at $T_C$ and $T_2$ are documented in Ref.[11], which supports MD anomaly around these two temperatures. A change in O-Nd-O bond angle and distortion of $NdO_6$ octahedron is documented below ~ 18 K.[11] A sharp change is observed in lattice parameters ('a', 'b' and 'c') and for the O-Nd-O angle below 25 K, followed by a distinct decrease/increase in the value of these parameters with decreasing temperature. Below 15 K they become nearly constant.[11] The variation of magnetic ordering in the system $Ba_3MRu_2O_9$ (for $M=$ Y, Nd, and La) is also attributed to the lattice effect. Therefore, the lattice anomaly and magnetic interactions are correlated, i.e., the change in magnetism yields MD coupling via large spin-orbit coupling. However, the change in lattice parameter and magnetic interaction turns out to be more complex in applied external magnetic fields $H$. This is documented through the $H$-dependent dielectric analyses for different orientation of $E$ and $H$. Note, that the system behaves like a soft ferromagnet, for which large hysteresis loop and strong MD effects can be observed at low temperatures, both could be related to magnetic domain wall motion as well.

The MD is quite complex as a function of $T$, $E$ and $H$. Probably, two or more different mechanisms are controlling the coupling, which is discussed in this manuscript. The negative MD could arise from magnetoelastic coupling below and above magnetic ordering, whereas, strong positive MD coupling may arise from D-M interaction below $T_2$ and dominates over negative MD coupling. The well-known multiferroic system $RMn_2O_5$ exhibits complex MD coupling, where D-M interaction and magnetostriction both are responsible,[23] eventually, possible magnetic-field-induced ferroelectricity far above magnetic ordering is recently reported through another mechanism.[24] The complex MD behavior is observed in some other rare-earth based compounds as well, where dielectric behavior is quite intriguing in presence of $H$.[25] Our bulk investigation on this interesting system warrants more detailed microscopic and theoretical studies to explore the exact mechanism of MD coupling, especially in external magnetic fields.

In summary, we have evaluated detailed magnetic and magnetodielectric coupling properties of a 6H-perovskite system containing Ru ($4d$-orbital), which is a good insulator at low temperature due to presence of the rare-earth ion. Magnetodielectric coupling is demonstrated in this Ru-based system. Interestingly, there is a cross-over of magnetodielectric coupling below ~$T_2$, which is more profound for a particular orientation of $E//H$. Magnetic correlations and weak magnetodielectric coupling is present even above $T_C$. The positive and negative magnetodielectric coupling is governed via two different mechanisms which are competing as a function of $T$, $E$ and $H$. Though the magnetodielectric coupling is rather weak, our primary results yield a path designing insulating systems which contain both rare-earth and higher $d$-orbital to achieve strong magnetodielectric coupling.

*Tathamay Basu,[1,2,*] Alain Pautrat,[1] Vincent Hardy,[1] Alois Loidl,[2] Stephan Krohn,[2]*

[1]Laboratoire CRISMAT, UMR 6508 du CNRS et de l'Ensicaen, 6 Bd Marechal Juin, 14050 Caen, France
[2] Experimental Physics V, Center for Electronic Correlations and Magnetism, University of Augsburg, Universitätsstr. 2, D-86135 Augsburg, Germany
[*]tathamaybasu@gmail.com

**Dc Magnetization:**

The dc magnetic susceptibility ($\chi=M/H$) as a function of temperature ($T$) is shown in figure S1a and S1b in presence of different magnetic fields (1, 10, 30 and 50 kOe) for zero-field-cooled (ZFC) and field-cooled (FC) conditions. A ferromagnetic type ordering is observed at $T_C$=24 K followed by a divergence between ZFC-FC at further low temperature for magnetic field of 1 kOe (figure S1a). The feature and value of the susceptibility agrees with earlier report by Doi, et al.[10] Under application of a magnetic field of 1 kOe, ZFC curve shows a maximum around 15 K followed by a sharp fall and become constant below 10 K, whereas, FC curve slowly goes towards saturation at low temperature. The application of higher magnetic field decreases the divergence between ZFC-FC features (see for 10 kOe ZFC-FC curve in figure S1b). No ZFC-FC divergence is observed for very high magnetic field (say, 30 kOe). The ZFC-FC divergence could occurs as a result of anisotropic domain in a ferromagnetic system. The inverse susceptibility is plotted as a function of temperature from 2-300 K for magnetic field of 10 kOe in the inset of figure S1b. The linearity of $\chi^{-1}(T)$ deviates below ~90 K. The Curie-Weiss temperature ($\Theta_{CW}$) and effective magnetic moment ($\mu_{eff}$), obtained from a liner fitting in paramagnetic region (inset of figure S1b), are -88.8 K and 5.16 $\mu_B$, respectively, nearly agreeing with earlier literature.[10] The negative $\Theta_{CW}$ indicates the presence of antiferromagnetic interaction in this system. Interestingly, $\chi$ does not superimpose high above $T_C$, at-least up to 35 K for the magnetic field of 10 and 30 kOe, as shown in figure S1b. The inverse susceptibility from 15-43 K is depicted for different magnetic field (1, 10, 30, 50 kOe) in the inset of figure S1a, which clearly shows bifurcation of $\chi^{-1}$ below ~35 K for different $H$. This indicates presence of magnetic correlation above long-range magnetic ordering.

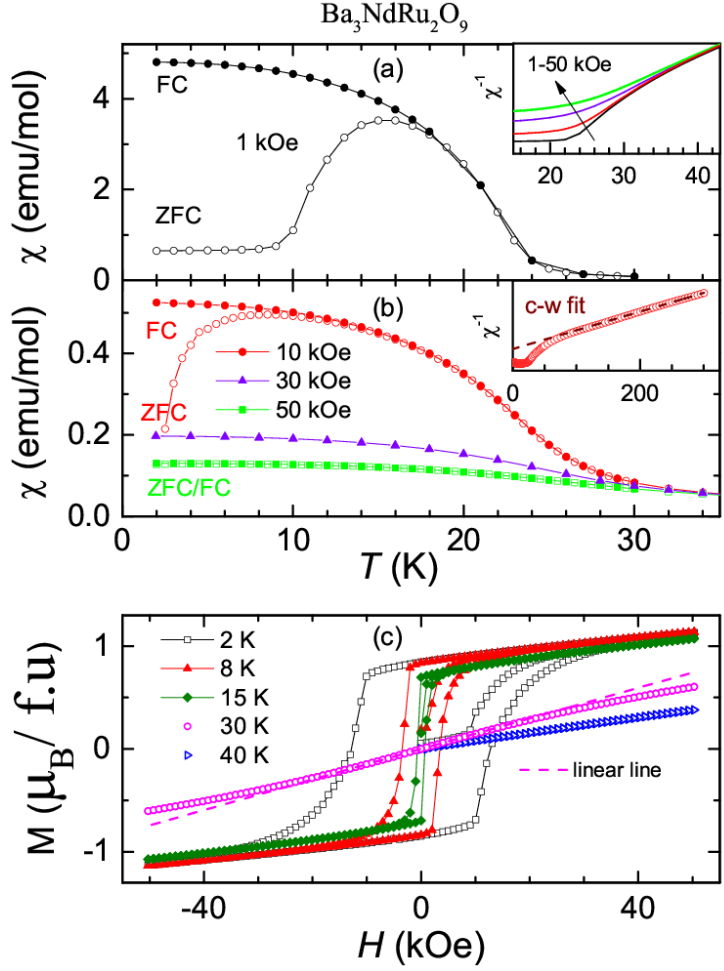

**Figure S1.** Dc magnetic susceptibility as a function of temperature from 0-35 K for different magnetic fields, (a) 1 kOe and (b) 10, 30 and 50 kOe), in zero-field-cooled and field-cooled conditions for the polycrystalline compound $Ba_3NdRu_2O_9$; Inset (a) : Inverse susceptibility as a function of temperature from 15-43 K for different magnetic field from 1-50 kOe (1, 10, 30 and 50 kOe); Inset (b) : Inverse susceptibility as a function of temperature from 2-300 K for magnetic field of 10 kOe and Curie-Weiss fit in paramagnetic region. (c) Isothermal dc magnetization as a function of magnetic field at fixed temperatures (2, 8, 15, 30 and 40 K); a linear line is drawn as a guide to the eye to compare the non-linearity of M(H) at 40 K.

Isothermal magnetization ($M(H)$) as a function of magnetic field is shown in figure S1c at several temperatures. The $M(H)$ results agrees with earlier report by Doi, et al.(Ref.[10]). Here, we have given a more detail explanations. Clear hysteresis loop below $T_C$ is observed, which confirms ferro(ferri)magnetic nature of this system. The coercive field at 2 K is quite large (10 kOe) and decreases with increasing temperature, though present at 15 K. This supports ZFC-FC divergence as a result of anisotropic ferromagnetic domain. The magnetization goes toward saturation at high magnetic field, though did not saturate completely even at very high field. Eventually, we have measured $M(H)$ up to 9 T employing a ACMS option in PPMS; instead of saturation, a slow linear increase is observed in $M(H)$ at 2 K (see figure S2); which is not a clear characteristic of a pure ferromagnetic system. These results are in well agreement with a canted antiferromagnetic nature at low temperature. The canting of the spin below 18 K for Nd-moments is predicted by neutron studies.[11] The $M(H)$ above $T_C$, say 30 K, does not exhibit a linear behavior; though $M(H)$ in paramagnetic region (i.e. Brillouin function at such high temperature) should be linear. This result documents the presence of magnetic correlation even above $T_C$ in this system, as inferred from *T*-dependent susceptibility. A linear dotted line is plotted with 30 K *M(H)* data as an guide to the eye to see the huge deviation from paramagnetic liner region at high magnetic field (figure S1c). The *M(H)* at 40 K exhibits almost linear behavior (figure S1c).

Finally, large ferromagnetic hysteresis loop without a complete saturation at very high magnetic field in isothermal *M(H)*, a ferromagnetic type ordering with a negative Curie-Weiss temperature in $\chi(T)$, presence of magnetic correlations above $T_C$ both in $\chi(T)$ and *M(H)*, infers a very complex magnetic interaction in this system. Magnetic frustration and canted antiferromagnetic behavior, as predicted by Senn, et al,[11] through neutron diffraction, is well supported by this results.

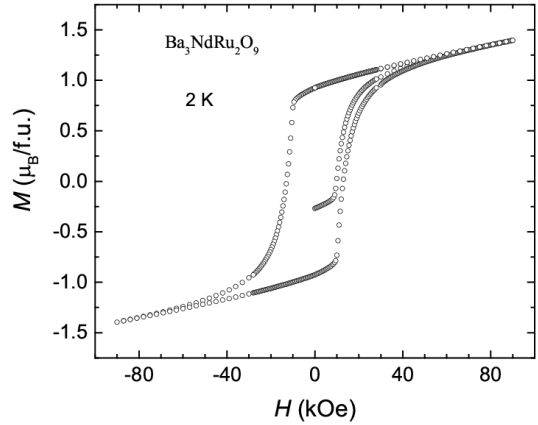

**Figure S2. Isothermal dc magnetization as a function of magnetic field at 2 K up to ±90 kOe (6 quad) for the polycrystalline compound Ba$_3$NdRu$_2$O$_9$.**

**Ac Magnetization:**

The real ($\chi'$) and imaginary ($\chi''$) part of ac susceptibility as a function of temperature for a different frequencies (1 Hz - 1.3 kHz) is shown in figure S3. No frequency dependence behavior is observed within the resolution limit of SQUID magnetometer for this compound.

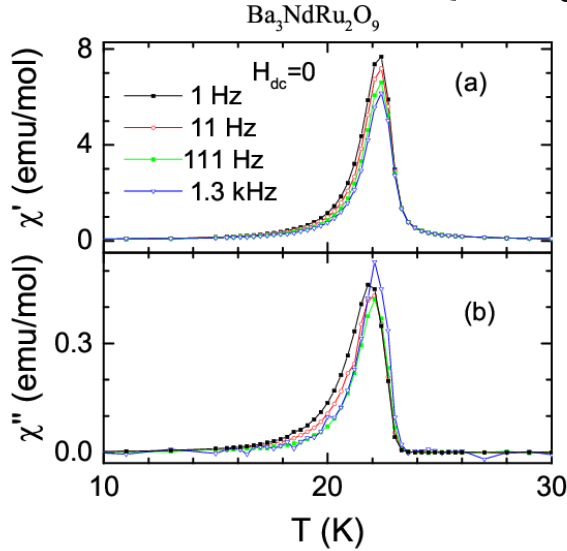

**Figure S3. (a) Real and (b) imaginary part of ac magnetic susceptibility as a function of temperature from 10-30 K for different frequencies (1Hz - 1.3 kHz) applying an ac magnetic field of 1 Oe in absence of dc magnetic field for the polycrystalline compound Ba$_3$NdRu$_2$O$_9$.**

**Dielectric constant:**

The real and imaginary part of dielectric constant is depicted in figure S4 from 7-42 K for different frequencies. The very low value of tanδ in magnetic regime reveals highly insulating nature of the sample to investigate MD behavior, which is discussed in the manuscript. The inset of figure S4 shows the *T*-dependent dielectric behavior from 7-120 K. A strong frequency dependent dielectric constant is observed around 70-90 K for the frequency range 11-82 kHz, which may arise as a result of hopping conductivity or from leakage current due to Maxwell–Wagner effect. There is no change in the peak temperature at the onset of magnetic ordering; however, there is a very small change in the absolute value of dielectric constant for different frequencies. The small change in the value could be due to overlap of left flange of high temperature dielectric features. We have not discussed the detail high temperature feature which is not the motivation of this manuscript.

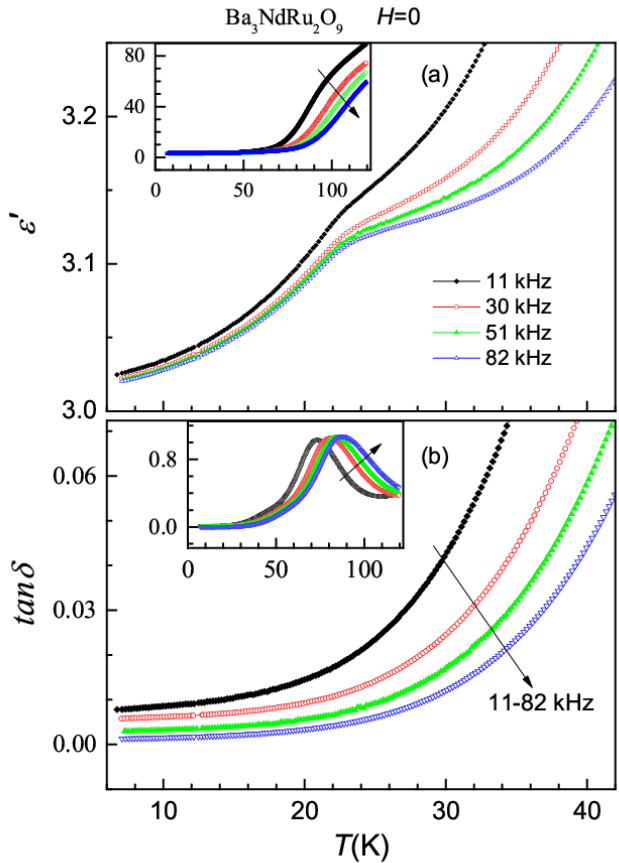

**Figure S4. (a) Real part of dielectric constant and (b) loss part tanδ as a function of temperature from 7-42 K in presence of different frequencies for the polycrystalline compound Ba$_3$NdRu$_2$O$_9$; Inset: shows the same for a wider temperature range from 7-120 K**